\documentclass{article}
\usepackage{fleqn,ams}
\usepackage{amssymb}
\usepackage{epsfig}
\usepackage{verbatim}
\newcommand{\const}{\mbox{Const}}
\def\stackunder#1#2{\mathrel{\mathop{#2}\limits_{#1}}}

\arraycolsep=\mathsurround
\newcommand{\Per}[1]{#1_{\perp}}

\begin{document}

\begin{center}
{\bf\large Exact  plane - symmetric non-stationary solution\\[2pt] of
the self-consistent equations of Einstein - Maxwell\\[2pt] for  a
magnetoactive plasma}\\[12pt]
Yu.G.Ignatyev, E.G.Chepkunova\\
Kazan State Pedagogical University,\\ Mezhlauk str., 1, Kazan
420021, Russia
\end{center}

\begin{abstract}
An exact plane-symmetric non-stationary solution to the
Einstein-Maxwell equations for a magnetoactive plasma is obtained
and studied.\end{abstract}

\section{Introduction}
In Ref. \cite{gmsw} on the basis of an exact solution to the
equations of relativistic magnetohydrodynamics (RMHD) against the
plane gravitational wave (PGW) background metric a new class of
relativistic essentially nonlinear phenomena arising in highly
magnetized plasma under the influence of the PGW was found and
called gravimagnetic shock waves (GMSW). The essence of the GMSW
phenomena consists in that the highly magnetized plasma:

\begin{equation} \label{1}
\alpha^{2}= \frac{\Per{H}^{2}}{4\pi(\varepsilon_{0} + p_{0})} \gg
1\,,
\end{equation}
(where $\Per{H}$ is the magnetic field intensity component which
is perpendicular in relation to the direction of the PGW
propagation, $\varepsilon_0$, $p_0$ - are the unperturbed energy
density and plasma pressure without taking a magnetic field into
account) anomalously highly responses even to a weak PGW at
sufficiently large values of the GMSW second parameter:
\begin{equation} \label{2}
\Upsilon \equiv 2 \beta_0 \alpha^2 > 1\,,
\end{equation}
where $\beta_0$ - the maximal amplitude of  the PGW.

In Ref.\cite{gmsw2} on the basis of the plasma and the PGW energy
balance model it is shown that the PGW is practically completely
transformed to the magnetoactive plasma acceleration (mainly in
the PGW propagation direction) and to the creating of a shock wave
with high densities of plasma and the magnetic field energy In the
mentioned works the isotropic plasma was considered locally, the
anisotropy was formed  exclusively by the magnetic field.

In the high magnetic fields as a result of magneto-bremsstrahlung
local thermodynamic equilibrium (LTE) in plasma is broken. The
given paper is devoted to a exact examination of this problem.

\section{Ricci tensor }

Let's  find solutions of the Einstein equations with a plane
symmetry when the symmetry "plane" is $\Pi \{ x^2 , x^3 \}$. The
metric of the space $V_4$ of the signature $(- 1, -1, -1, + 1)$,
supposing two spacelike vectors of Killing:

\begin{equation}
\label{2.1} \stackunder{1}{\xi}^i = \delta^i_2\,; \quad
\stackunder{2}{i} = \delta^i_3
\end{equation}

and the corresponding PGW symmetry with the polarization $e_+$ can
be written in the form:

\begin{equation}  \label{2.2}
ds^2 = \Phi - L^2 \left[ e^{2 \beta} (d x^2)^2 + e^{- 2 \beta} (dx^3)^2
\right]\,,
\end{equation}

where

\begin{equation}  \label{2.3}
L = L(x^1, x^4)\,; \quad \beta = \beta (x^1, x^4)
\end{equation}
and
\begin{equation}   \label{2.4}
\Phi = d \sigma^2 = g_{\alpha \beta}(x^1, x^4)\; \quad (\alpha ,
\beta = \overline{1, 2})
\end{equation}
- is the metric of the two-dimensional pseudoeuclidean surface
$\Sigma$: $x^2 = \const$; $x^3 = \const$, and:
\begin{equation}   \label{2.5}
g_{\alpha \beta} = g_{\alpha \beta}(x^1, x^4)\,.
\end{equation}
Thus,
\begin{equation}  \label{2.6}
V_4 = \Sigma \times \Pi_{\Sigma}\,.
\end{equation}

As it is known (for example see Ref. \cite{nord}), the
two-dimensional surface metric can be always reduced to a
conform-plane form:
\begin{equation}   \label{2.7}
\Phi = e^{2 \lambda} \left[ (dx^4)^2 - (dx^1)^2\right]\,.
\end{equation}
Thus, with the help of the admissible transformations of the
coordinates

\begin{equation}  \label{2.8} x^{1'} = f^1 (x^1,
x^4)\,; \quad x^{4'} = f^4 (x^1, x^4)\,,
\end{equation}
which do not alter the metric
$\Pi_{\Sigma}$, the metric $V_4$ can be reduced to

$$
ds^2 = e^{2 \lambda} \left[ (dx^4)^2 - (dx^1)^2 \right] -
$$
\begin{equation}  \label{2.9}
- L^2 \left[ e^{2 \beta} (dx^2)^2 + e^{- 2 \beta} (dx^3)^2 \right]\,,
\end{equation}
where
$$
\lambda = \lambda (x^1, x^4)\,.
$$

Note, the metric of Eq.(\ref{2.9}) is invariant in relation to
Lorentz transformations in the plane $\Sigma$. In the retarded,
$u$, and advanced, $v$, time coordinates:
$$  \displaystyle
u = \frac{1}{\sqrt{2}} (t - x)\, \quad v = \frac{1}{\sqrt{2}} (t +
x)\,,
$$
The metric of Eq. (\ref{2.9}) takes the form:
\begin{equation}  \label{2.14}
ds^2 = 2 e^{2\lambda} du dv - L^2 \left( e^{2\beta}(dx^2)^2 +
e^{- 2\beta}(dx^3)^2 \right)\,.
\end{equation}
The metric of Eq. (\ref{2.9}) coincides with the rotationally
symmetric metric, see Ref. \cite{sing}. Hence, we obtain Ricci
tensor nonzero components which in the coordinates $u$ and $v$
take the form:
\begin{equation}  \label{2.15}  \displaystyle
R_{uu} = - \frac{2}{L} (L_{uu} + L \beta^2_u  - 2\lambda_u L_u )\,;
\end{equation}
\begin{equation}  \label{2.16}  \displaystyle
R_{vv} = - \frac{2}{L} (L_{vv} + L \beta^2_v - 2 \lambda_v L_v
)\,;
\end{equation}
\begin{equation}  \label{2.17}  \displaystyle
R_{uv} = - \frac{2}{L} (L_{uv} + L \beta_u \beta_v + L \lambda_{uv} )\,;
\end{equation}
\begin{equation}  \label{2.18}  \displaystyle
R^2_2 + R^3_3 = - 2 e^{- 2\lambda} \frac{(L^2)_{uv}}{L^2}\,;
\end{equation}
\begin{equation}  \label{2.19} \displaystyle
R^2_2 - R^3_3 =  - 4 e^{- 2\lambda} \left( \beta_{uv} + \frac{L_u
\beta_v + L_v \beta_u }{L} \right)\,.
\end{equation}

\section{Conditions of a magnetic field freezing-in into plasma and the Einstein's equations}
Let's study magnetoactive plasma moving in the metric
(\ref{2.14}). In Ref.\cite{gmsw} it is shown that the
electromagnetic field of the magnetoactive plasma should satisfy
freezing-in in plasma:
\begin{equation}  \label{3.1}
F_{ik} v^k = 0\,,
\end{equation}
where $v^k$ is the vector of the plasma dynamic velocity that
coincides with the vector of the electromagnetic field dynamic
velocity (by Synge, Ref.\cite{sing}). Thus, the first invariant of
the electromagnetic field equals to null:
\begin{equation}  \label{3.2}
F^{ik} \stackrel{*}{F}_{ik} = 0\,,
\end{equation}
and the second invariant is positive:
\begin{equation}  \label{3.3}
\frac{1}{2} F_{ik} F^{ik} = - (H, H) \equiv H^2 > 0\,,
\end{equation}
- where  $H_i$ is the magnetic field vector:
\begin{equation}  \label{3.4}
H_i = v^k \stackrel{*}{F}_{ki}\,,
\end{equation}
and $\stackrel{*}{F}_{ki}$ is the tensor which is dual to the
Maxwell antisymmetric tensor $F_{ki}$.

In Ref.\cite{gmsw} it is shown that the complete tensor of an
impulse energy (EIT) of locally isotropic magnetoactive plasma
takes the form:
\begin{equation}  \label{3.5}
T_{ik} = ({\cal E} + P) v_{i} v_{k} - P g_{ik} -
2 P_{H} n_{i} n_{k},
\end{equation}
where
\begin{equation} \label{3.6}
P_H = \frac{H^2}{8 \pi}; \quad
{\cal E} = \varepsilon + \varepsilon_H \,;
\quad P = p + P_H,
\end{equation}
$P,{\cal E}$ - are the summary pressure and density of the
magnetoactive plasma energy, and

\begin{equation} \label{3.7} n_i
= \frac{H_i}{H} -
\end{equation}

is the single spacelike vector of the magnetic field:
\begin{equation}  \label{3.8}
(n,n) = -1,
\end{equation}
where
\begin{equation}  \label{3.9}
(n,v) = 0.
\end{equation}

Under the conditions of a plane symmetry we shall consider the
plasma spreading in the direction of $x^1$ and the magnetic field
directed along $x^2$. The vector potential corresponds to this
field, see Ref.\cite{gmsw}:
\begin{equation}  \label{3.10}
A_u = A_v = A_2 = 0\,; \quad A_3 = \psi (u,v)\,,
\end{equation}
where $\psi$ - is an arbitrary function of its arguments.
Calculating the Maxwell tensor in relation to the potential
(\ref{3.10}) we shall find its nonzero components:
\begin{equation}  \label{3.11}
F_{u3} = \psi_u\,; \quad F_v3 = \psi_v\,.
\end{equation}
Calculating the invariant (\ref{3.3}) we shall obtain:
\begin{equation}  \label{3.12}  \displaystyle
H^2 = - 2 \frac{e^{2\beta} \psi_u \psi_v}{L^2 e^{2 \lambda}}\,.
\end{equation}
Thus, it should be $\psi_u \psi_v < 0$ . Let us choose:
\begin{equation}  \label{3.13}
\psi_u < 0\,; \quad \psi_v > 0\,,
\end{equation}
the magnetic field positive direction corresponds to this choice:
$n^2 > 0$.

Then taking the relation of the velocity vector normalization into
consideration the conditions of freezing-in (\ref{3.1}) for the
nonzero components of this vector give:
\begin{equation}  \label{3.14}  \displaystyle
v_v = e^{\lambda} \sqrt{- \frac{\psi_v}{\psi_u}}\,; \quad
v_u = e^{\lambda} \sqrt{- \frac{\psi_u}{\psi_v}}\,.
\end{equation}
Taking these relations and Eq.(\ref{3.5}) into account let us
write Einstein's nontrivial equations out:

\begin{equation}  \label{3.16}  \displaystyle
 L_{uu} + L \beta^2_u - 2 \lambda_u L_u  = \varkappa \frac{L e^{2\lambda}
\psi_u}{4 \psi_v} \left( \varepsilon + p + \frac{H^2}{4 \pi} \right)\,;
\end{equation}
\begin{equation}  \label{3.17}  \displaystyle
 L_{vv} + L \beta^2_v - 2 \lambda_v L_v  = \varkappa \frac{L e^{2\lambda}
\psi_v}{4 \psi_u} \left( \varepsilon + p + \frac{H^2}{4 \pi} \right)\,;
\end{equation}
\begin{equation} \label{3.18}
L_{uv} + L \beta_u \beta_v + L \lambda_{uv} = -
\frac{\varkappa}{2} L e^{2 \lambda} p\,;
\end{equation}
\begin{equation}  \label{3.19}
(L^2)_{uv} = \frac{\varkappa}{2} L^2 e^{2 \lambda} (\varepsilon -
p)\,;
\end{equation}
\begin{equation}  \label{3.20}  \displaystyle
\beta_{uv} + \frac{L_u \beta_v + L_v \beta_u}{L} = -
\frac{\varkappa}{16\pi} e^{2\lambda} H^2\,.
\end{equation}

The Eqs.(\ref{3.16}) simultaneously with the definition $H^2$
(\ref{3.12}) and the local equation of the plasma state:
\begin{equation}  \label{3.21}
p = p(\varepsilon )
\end{equation}
represent a complete equations set in relation to the five
unknowns of the scalar functions: $\lambda$, $\beta$, $L$, $\psi$,
$\varepsilon$. Note dew to the last equation of the
Eqs.(\ref{3.20}):
\begin{equation}  \label{3.22}
\beta = \const \rightarrow H^2 = 0\,.
\end{equation}

It is this property that calls in question the correctness of the
energy balance model used in Refs.\cite{gmsw} and \cite{gmsw2} in
order to take the GMSW influence back on the gravitational wave
metric into account.

If we suppose all functions depending only on one variable $t$ in
Eqs.(\ref{3.16}) - (\ref{3.20}), we shall obtain a homogeneous
anisotropic universe model with a magnetic field. If we suppose
all functions depending only on the variable $x$, we shall obtain
a static model of the plane anisotropic stratum. In vacuum the
system (\ref{3.16}) - (\ref{3.20}) supposes also a retarded
solution (all functions depend only on the variable $u$), or an
advanced solution (all functions depend only on the variable $v$),
called plane gravitational waves. In these cases from the system
(\ref{3.16}) - (\ref{3.20}) one nontrivial equation for three
metric functions is left. It gives an opportunity to choose, for
example:
\begin{equation}  \label{gauge}
\lambda (u) = 0\,,
\end{equation}

and to let the function $\beta (u)$ which is the PGW amplitude be
arbitrary. Then for the PGW background factor, i.e. for $L(u)$, we
get the equation (see Ref. \cite{torn})
\begin{equation}
\label{pgw} L_{uu} + \beta^2_u L = 0\,.
\end{equation}

\section{Static solution}
Let us assume there is no gravitational wave
\begin{equation}  \label{4.1}
\psi = \psi (x)\,,
\end{equation},
then we get $\psi_u = - \psi_v$, and according to the
Eq.(\ref{3.14}) $v_u = v_v$ $\rightarrow v^1 =0$, i.e. plasma is
at rest. For a static metric the first two equations of the
Eqs.(\ref{3.16}) è (\ref{3.17}) coincide, the independent
Einstein's equations take the form:
\begin{equation}  \label{4.2} \displaystyle
L'' + L \beta'^2 - 2 \lambda' L' = - \frac{\varkappa}{2} L e^{2
\lambda} \left( \varepsilon + p + \frac{H^2}{4 \pi}\right)\,;
\end{equation}
\begin{equation}   \label{4.3}
L'' + L \beta'^2 +  \lambda'' L = \varkappa L e^{2 \lambda} p\,;
\end{equation}
\begin{equation}  \label{4.4}
(L^2)'' = - \varkappa L^2 e^{2 \lambda} ( \varepsilon - p)\,;
\end{equation}
\begin{equation}  \label{4.5}  \displaystyle
\beta'' + 2 \beta' \frac{L'}{L} =
\frac{\varkappa}{16\pi}e^{2\lambda}H^2\,.
\end{equation}
From the four equations Eqs.(\ref{4.2}) - (\ref{4.5}) two
equations Eqs.(\ref{4.4}) and Eqs.(\ref{4.5}) are the definitions
of $\varepsilon$ and $H^2$. Thus, for the three metric functions
$\lambda$, $\beta$ and $L$ there are only two equations, it gives
an opportunity to impose  an additional condition  on  these
functions, defining a class of solutions.

From the set (\ref{4.6}), having carried out identical
transformations, we get a consequence:
\begin{equation}  \label{4.7}  \displaystyle
L'' + L\beta'^2 + L\lambda'' = - \frac{1}{2L}[L^2 \nu' - \frac{1}{2}
(L^2)'\, ]'\,,
\end{equation}
introducing a new variable:
\begin{equation}  \label{4.8}
\nu = \beta - \lambda \,.
\end{equation}

we write the consequence in the form
\begin{equation}
L''+L {\beta '}^2+L \lambda''=-\frac{1}{2L} [L^2 \nu ' -
\frac{1}{2}(L^2)']'
\end{equation}

With the help of the Eq.(\ref{4.7}) the relations (\ref{4.3}) -
(\ref{4.6}) can be reduced to a more convenient form:
\begin{equation} \label{4.9} \displaystyle
[L^2 \nu' - \frac{1}{2}(L^2)'\, ]' = - 2\varkappa L^2 e^{2\lambda}
p\,;
\end{equation}
\begin{equation}  \label{4.10}
(L^2)'' = - \varkappa L^2 e^{2\lambda}(\varepsilon - p)\,;
\end{equation}
\begin{equation}  \label{4.11}  \displaystyle
(\beta' L^2 )' = - \frac{\varkappa}{16\pi} L^2 e^{2\lambda} H^2\,.
\end{equation}

If we accept the barotropic equation
\begin{equation}  \label{4.12}
p = k \varepsilon\,; \quad (0 < k < 1)\,.
\end{equation}
as a local equation of the plasma state;

From the Eqs.(\ref{4.9}) and (\ref{4.10}) we shall obtain an
algebraic consequence:
\begin{equation}   \label{4.13} \displaystyle
[(1 - k)L^2 \nu' - \frac{1+3k}{2}(L^2)'\,]' = 0\,,
\end{equation}
from which we get the first integral:
\begin{equation}  \label{4.14} \displaystyle
L^2\nu' = C_1 + \frac{1+3k}{2(1-k)}(L^2)'\,,
\end{equation}
where $C_1 = \const$. Dew to Eq.(\ref{4.14}) the following
relations are valid, for example:
$$  \displaystyle
[L^2 \nu' - \frac{1}{2}(L^2)'\, ]' = \frac{2k}{1-k}(L^2)''\,;
$$
$$  \displaystyle
[L^2 \nu' + \frac{1}{2}(L^2)'\, ]' = \frac{1+k}{1-k}(L^2)''\,.
$$

With the help of them the definitions of the plasma energy density
(\ref{4.10}) and the  magnetic field (\ref{4.11}) can be written
in a more symmetric form:
\begin{equation}  \label{4.15}  \displaystyle
\varkappa \varepsilon = - \frac{2 e^{- 2\lambda}}{1+3k}\frac{(L^2
\nu'\,)'}{L^2} \,;
\end{equation}
\begin{equation}  \label{4.16}  \displaystyle
\varkappa \frac{H^2}{8\pi} = - \frac{2 e^{- 2\lambda}(L^2
\beta'\,)'}{L^2} \,.
\end{equation}
Dew to the nonnegativity of the functions $\varepsilon$ and $H_2$
the solution of the Einstein equations should satisfy the
conditions:
\begin{equation}  \label{4.16a}
(L^2 \nu'\,)' \leq 0\,; \quad (L^2 \beta'\,)' \leq 0 \,.
\end{equation}

The use of the integral (\ref{4.14}) allows us to write the first
of the conditions (\ref{4.16a}) in a more compact form:
\begin{equation}  \label{4.14b}
(L^2)'' \leq 0\,.
\end{equation}

Thus we obtain the following local value of the parameter
$\alpha^2 (x)$ introduced in Ref.\cite{gmsw}:
\begin{equation}  \label{4.17}  \displaystyle
\alpha^2 = \frac{H^2}{4\pi (\varepsilon + p)} = 2 \frac{(1+3k)}{(1+k)}
\frac{(L^2 \beta')\,'}{(L^2 \nu')\,'}\,.
\end{equation}

In the case of the barotropic equation of state the Einstein's set
of equations is reduced to two independent nonlinear differential
equations for the three metric functions, one of which,
Eq.(\ref{4.14}), is of the first order and the second one,
Eq.(\ref{4.7}), is of the second order. It is possible to impose
one additional condition on the three functions which does not
contradict Eq.(\ref{4.16a}). Choosing in the integral
(\ref{4.14}):
\begin{equation} \label{4.18}
C_1 = 0\,,
\end{equation}
we get a private solution:
\begin{equation}  \label{4.19} \displaystyle
\nu =  \frac{1 + 3k}{1 - k} \ln L\,.
\end{equation}
Being free in choosing the additional condition for the metric
functions, we can assume, for example, in Eq.(\ref{4.17}):
\begin{equation}  \label{4.20}
\alpha^2 = \alpha^2_0 = \const \,.
\end{equation}
Assuming $\alpha^2_0 \neq 0$, from Eq.(\ref{4.17}) we get one more
first integral:
\begin{equation}  \label{4.21}  \displaystyle
L^2 \nu' =  \frac{2(1+3k)}{\alpha^2_0 (1+k)} L^2 \beta' + C_2 \,,
\end{equation}
where $C_2 = \const$.

Choosing here again :
\begin{equation}  \label{4.22}
C_2 = 0\,,
\end{equation}
we have:
\begin{equation}  \label{4.23}  \displaystyle
\beta = \alpha^2_0 \frac{1 + k}{2(1 - k)} \ln L \,.
\end{equation}
Let us write the integrals (\ref{4.19}) and (\ref{4.23}) in the
form:
$$
\nu =  q_1  \ln L\,  ,
$$
where
$$
q_1=q_1(k) =\frac{1 + 3k}{1 - k}
$$
$$
\beta = q_2  \ln L \, ,
$$
where
$$
q_2=q_2 (k, \alpha_0)=\alpha^2_0 \frac{1 + k}{2(1 - k)}.
$$

Substituting these integrals in the left Eq.(\ref{4.7}) we obtain
the equation closed in relation to the function $L$:
\begin{equation}   \label{4.24}  \displaystyle
L'' L = q_3 L'^2,
\end{equation}
where
$$
q_3=q_3 (k, \alpha_0)=\frac{1-3q_1+2q_2-2q_2^2}{1+2q_2-q_1}.
$$

Solving it we get:
\begin{equation}   \label{4.25}  \displaystyle
L = \left(\mu_1(v) u (1+q_3) +\mu_2(v)\right)^{\displaystyle
\frac{1}{1+q_3}}
\end{equation}
where $\mu_1(v)$, $\mu_2(v)$ -- are arbitrary functions. Let us
define their form from the static condition of solution:
$$\psi=\psi(x)=\psi(v-u).$$

whence it follows:
\begin{equation}\label{4.26} \mu_1(v) =
-\frac{A}{\sqrt{2}(1+q_3)}=\const;
\end{equation}
\begin{equation}
\mu_2(v) = \frac{A}{\sqrt{2}}\, v +B.
\end{equation}

Coming from the variable $v$ to the variable $x$, according to
Eq.(2), expression (\ref{4.25}) takes the form:
\begin{equation}\label{4.27}
L = \left( A x + B \right)^{\displaystyle q_4}
\end{equation}
where
\begin{equation}
q_4=q_4 (k, \alpha_0)=\frac{1}{1+q_3},
\end{equation}
Supposing $B=1$ we obtain a condition on the function  $L$:
\begin{equation}
L(0)=1
\end{equation}

And the solution (\ref{4.27}) takes the form:
\begin{equation}\label{4.27a}
L = \left( A x + 1 \right)^{\displaystyle q_4}
\end{equation}

Substituting the obtained solution into Eqs. (\ref{4.8}),
(\ref{4.16}) and (\ref{4.23}) we have

\begin{equation}\label{4.30}
\lambda =  q_4 (q_2-q_1)\ln \left( A x + 1 \right)
\end{equation}

\begin{equation} \label{4.28a}
\varepsilon=\varepsilon_0 (A x + 1)^{\displaystyle q_5 },
\end{equation}
where
$$
\varepsilon_0 =   \frac{2 A^2 q_1 q_4 (1-2q_4)} {\varkappa
(1+3k)},
$$
$$
q_5 = q_5 (k, \alpha_0)=-2 q_4 (q_2-q_1)-2.
$$
\begin{equation}   \label{4.29}  \displaystyle
H^2=H^2_0 (A x +1)^{\displaystyle q_5}
\end{equation}
where
$$
H^2_0=\frac{16 \pi A^2 q_2 q_4 (1-2 q_4)}{\varkappa}.
$$
\begin{equation}\label{4.28}\displaystyle
\nu =  q_1 q_4 \ln \left( A x + 1 \right)
\end{equation}
\begin{equation}\label{4.31}
\beta = q_2 q_4 \ln \left( A x + 1 \right)
\end{equation}

Thus the metric (\ref{2.9}) is
\begin{eqnarray}  \label{4.32}
&  ds^2 = \left(A x^1+1 \right)^{ \displaystyle 2 q_4(q_2-q_1)}
\left[ (dx^4)^2 - (dx^1)^2
\right] - \nonumber \\
& - \Bigl[\displaystyle \left( A x^1 + 1 \right)^
{\displaystyle 2q_4(1+ q_2)} (dx^2)^2 +\nonumber \\
& + \left( A x^1 + 1 \right)^ {\displaystyle 2q_4(1-q_2)} (dx^3)^2
\Bigr]\,,
\end{eqnarray}

Constant $A$ can be easily derived from Eq.(\ref{4.29}):
\begin{equation}
A^2=\frac{H^2_0 \varkappa}{16 \pi q_2 q_4 (1-2 q_4)}
\end{equation}

For the metric (\ref{4.32}) let us obtain the following nonzero
components of the Riemann's tensor:
\begin{equation}
R_{2323}=-{q_4}^{2}\left ({q_2}^2-1\right ){A}^{2}\left (A
x^1+1\right )^{\displaystyle q_5+4q_4},
\end{equation}
\begin{equation}
R_{3434}={q_4} ^{2}\left (q_2-q_1\right )\left (q_2-1\right
){A}^{2}\left (A x^1+1 \right )^{\displaystyle -2-2\,q_4(q_2-1)},
\end{equation}
$$
R_{1313}= q_4\,\left (q_2-1\right ){A}^{ 2}\left (q_4(2\,
q_2-q_1-1)+\right.
$$
\begin{equation}
\left.+1 \right )\left (A x^1+1\right )^{\displaystyle
-2\,q_4(q_2-1)-2},
\end{equation}
$$
R_{1212}= q_4\,\left (1+q_2\right ){A}^{2} \left
(q_4+q_4\,q_1-\right.
$$
\begin{equation}
\left. -1\right )\left (A x^1+1\right )^{\displaystyle 2
q_4(q_2+1)-2},
\end{equation}
$$
R_{2424}=-{q_4}^{2}\left (q_2-q_1\right ){A}^{2}\left (1+\right.
$$
\begin{equation}
\left.+q_2\right )\left (A x^1+ 1\right )^{\displaystyle 2\,
q_4(q_2+1)-2},
\end{equation}
\begin{equation}
R_{1414}=q_4\,\left (q_2-q_1\right ){A}^{2}\left (A x^1+1\right
)^{\displaystyle q_5}
\end{equation}

In the limit $\alpha_0 \rightarrow \infty$, from the mentioned
function it follows:
\begin{eqnarray}
q_2 = \infty \quad q_3=\infty\quad q_4=0\quad q_5=0
\end{eqnarray}

And the metric (\ref{4.32}) is reduced to:
\begin{eqnarray}
& ds^2=(A x^1 +1)^{-2} \left[(dx^4)^2-(dx^1)^2\right]-\nonumber\\
& -\left[ (A x^1 +1)^{-2}(dx^2)^2+(A x^1 +1)^2(dx^3)^2\right]
\end{eqnarray}

If $\alpha_0 \rightarrow 0$ we obtain
$$q_2 = 0$$
\begin{equation}
q_3=\frac{5k+1}{2k} \quad q_4=\frac{2k}{7k+1}
\end{equation}
$$q_5=-\frac{2(13k^2-4k-1)}{7k^2-6k-1}$$

Then the metric (\ref{4.32}) will take the form
\begin{eqnarray}  \label{4.32_0}
&  ds^2 = \left(A x^1+1 \right)^{ \displaystyle
\frac{4k(1+3k)}{(7k+1)(k-1)}} \left[ (dx^4)^2 - (dx^1)^2
\right] - \nonumber \\
& - \displaystyle \left( A x^1 + 1 \right)^ {\displaystyle
\frac{4k}{7k+1}} \Bigl[(dx^2)^2 +(dx^3)^2 \Bigr]\,,
\end{eqnarray}

Let us study some special cases. If $k = 0$ the metric
(\ref{4.32}) is transformed to
\begin{equation}
ds^2 =(dx^4)^2 - (dx^1)^2-(dx^2)^2 -(dx^3)^2
\end{equation}

Supposing $k = 1/3$, we have
\begin{eqnarray}  \label{4.32_0}
&  ds^2 = \left(A x^1+1 \right)^{ \displaystyle -\frac{6}{5}}
\left[ (dx^4)^2 - (dx^1)^2
\right] - \nonumber \\
& - \displaystyle \left( A x^1 + 1 \right)^ {\displaystyle
\frac{2}{5}} \Bigl[(dx^2)^2 +(dx^3)^2 \Bigr]\,,
\end{eqnarray}

For $k = 1$ we get
\begin{eqnarray}  \label{4.32_0}
&  ds^2 = \left(A x^1+1 \right)^{ \displaystyle \infty} \left[
(dx^4)^2 - (dx^1)^2
\right] - \nonumber \\
& - \displaystyle \left( A x^1 + 1 \right)^ {\displaystyle
\frac{1}{2}} \Bigl[(dx^2)^2 +(dx^3)^2 \Bigr]\,,
\end{eqnarray}

\end{document}